\documentstyle[12pt, aasms4, epsf, rotate]{article}

\def\ugr{\, \lower4pt \hbox{$\buildrel > \over \sim$} \, }
\def\ukl{\, \lower4pt \hbox{$\buildrel < \over \sim$} \, }

\lefthead{B\"ottcher \& Liang}
\righthead{Comptonization signatures in the rapid aperiodic variability
of GBHCs}

\begin{document}

\title{Comptonization signatures in the rapid aperiodic variability of 
Galactic Black-Hole Candidates}
\author{M.  B\"ottcher, E. P. Liang \altaffilmark{1}}
\altaffiltext{1}{Rice University, Space Physics and Astronomy Department, MS 108\\
6100 S. Main Street, Houston, TX 77005 -- 1892, USA}

\begin{abstract}
We investigate the effect of inverse-Compton scattering 
of flares of soft radiation in different geometries 
of a hot, Comptonizing region and a colder accretion 
disk around a solar-mass black hole. The photon-energy 
dependent light curves, their Fourier transforms, 
power spectra and Fourier-period dependent time 
lags of hard photons with respect to softer photons 
are discussed. On the basis of a comparison with existing
data we find arguments against Comptonization of external
soft radiation as well as Comptonization in a homogeneous
medium as dominant mechanisms for the rapid aperiodic
variability in Galactic black-hole candidates. Possible 
further observational tests for the influence of Comptonization
on the rapid aperiodic variability of Galactic black-hole
candidates are suggested.
\end{abstract}

\keywords{X-rays: stars --- accretion, accretion disks ---
black hole physics --- radiative transfer --- 
radiation mechanisms: thermal}

\section{Introduction}

The rapid aperiodic and quasiperiodic variability of the
X-ray signals from Galactic black-hole candidates (GBHCs)
and low-mass X-ray binaries (for a review see \cite{vdk95}) 
contains valuable information about the source of high-energy 
emission in these objects. Its typical timescales, typical 
repetition frequency, power spectrum density (PSD), autocorrelation 
function, time lags between different energy bands etc. give 
hints toward important parameters such as the extent of 
the emitting region, the dominant microscopic timescales,
the dominant emission mechanism and the geometry of the source.

With the new generation of X-ray telescopes such as the PCA
and the HEXTE on board RXTE it is now possible to measure
the above properties of the rapid variability of X-ray
binaries in great detail with high timing resolution 
and good spectral resolution. As we will see in this 
paper, the photon-energy dependence of the rapid 
aperiodic variability can provide important diagnostics 
of the nature of the X-ray emitting  regions. 

Early measurements of the Fourier-frequency dependence
of time lags between the signals in different X-ray
energy channels from GBHCs (\cite{mkk88}, \cite{mik93}) 
have been interpreted as evidence that Comptonization of 
soft photons in a hot, uniform plasma could not be the 
dominating mechanism for
the production of hard X-rays. This conslusion, however, 
is strongly geometry-dependent and does not hold for a very 
extended, inhomogeneous Comptonizing region, as was recently
found by Kazanas et al. (1997). In that and in two
subsequent papers (\cite{hkt97a}, \cite{hkc97b}) the
effect of radial density gradients in the hot Comptonizing
regions in GBHCs on the power density spectra and the
phase and time lags between different energy bands
were discussed. From the comparison of the measured 
Fourier-frequency dependent time lags between two
energy channels of the signal from Cyg~X-1, they
deduced that a radial density dependence $n(r) \propto
r^{-1}$ in the Comptonizing hot plasma is appropriate 
to account for the observed hard time lags and the hard 
X-ray spectrum at the same time.

However, their work was restricted to a geometry with a
central soft photon source surrounded by a spherical 
corona of hot Comptonizing plasma. Furthermore, the 
density was the only parameter allowed to vary radially.

In this paper, we extend the investigation of the effects
of Comptonization the rapid variability in X-ray binaries, 
discussing two fundamentally different source geometries 
and radial gradients of other physical parameters, such 
as the electron temperature. This is done primarily 
with Monte-Carlo simulations of instantaneous
flares of soft radiation (described as a $\delta$-function
in time), being Comptonized by the X-ray emitting region. 
The generalization to the case of multiple, randomly 
distributed flares (shot-noise) is straightforward.
The light curves resulting from the Monte-Carlo simulations
are then Fourier transformed, and the PSD as well as the 
Fourier-frequency dependent phase and time lags between 
different energy bands are calculated.

In Section 2, we describe our Monte-Carlo simulations
leading to the energy-dependent light curves for the
different geometrical situations, together with analytical 
approximations for these light curves. Their Fourier 
transforms and the corresponding time lags between different 
energy bands will be discussed in Section 3. In Section 4, 
we investigate specific differences between the PSDs 
and hard time lags resulting from different geometries
and compare the predictions of both scenarios to the power
spectra and hard time lags observed for some GBHCs. We will 
suggest further observational tests which could either
confirm a preferred geometry, or rule out to the Comptonization 
scenario in general.

\section{Simulated light curves}

For our time-dependent Monte-Carlo simulations on Compton
scattering in a hot plasma we use an improved version of
the Comptonization Code developed by Canfield et al. (1987), 
see also Liang (1993), Wandel \& Liang (1991).
It can handle spherically symmetric or slab geometry of the
Comptonizing region with thermal soft photon sources located
in the center of the Comptonizing cloud, distributed
throughout the cloud, and/or located outside the Comptonizing
region. All parameters (such as density, electron temperature,
radiation temperature of an internal soft photon field) may
have arbitrary radial dependencies within the cloud. 

Practically, we split the Comptonizing region into 20 radial 
zones, each zone having the same (radial) Thomson depth, within 
which the parameters are set to a constant value corresponding 
to an average of the respective radial dependencies over the
zone. For several of our simulations, we run test simulations 
with a finer radial grid (50 -- 100 zones) in order to check 
that the step function dependency of the parameters in the 
Comptonizing region does not introduce artificial structures 
in the light curves. No significant difference to our simpler
(and faster) 20 zone simulations has been found in any of
these tests.

In this paper, we focus on two basic situations with respect
to the location of the flaring soft photon source: a source
located at the center of the cloud and a source located
outside the cloud. The former situation is representative 
of an accretion-disk-corona model (\cite{lp77}, \cite{bk77}, 
\cite{hm93}) where photons resulting from a flare in the cold, 
optically thick disk are Compton-upscattered by the hot corona to
produce the hard X-ray spectrum. The latter case represents 
a hot-inner-disk model (\cite{sle76}) 
or an advection-dominated accretion flow (\cite{ny94}, 
\cite{cal95}), assuming that a soft photon flare occurs 
in the cold, optically thick outer disk and its radiation 
impinges on the hot inner disk (torus), producing the hard 
X-rays via Compton upscattering.

For all simulations shown in this paper, the soft photon 
spectrum is assumed to be a thermal blackbody of $kT_r = 0.2$~keV.

For each of the light curves resulting from central soft 
photon injection, a total of $9 \cdot 10^6$ photons, being 
injected instantaneously at $t = 0$, are simulated. The photons 
leaving the system are collected in 5 energy channels and in 
equal time steps appropriate to reveal the detailed structure 
of the light curves (typically $\Delta t \sim 10^{-2} \, R/c$,
where $R$ is the radius of the Comptonizing region). The
light curves resulting from external soft photon injection
turn out to be mathematically more complicated and require a 
higher numerical signal-to-noise ratio in order to yield
reliable Fourier transforms. Therefore, the simulations
for this geometry have been carried out with 25~million
photons each.

For both geometries, we investigate the effect of density
and temperature gradients within the hot Comptonizing region.
Fig. 1 shows the energy-dependent light curves for the 
case of a central photon source inside a homogeneous
cloud. We did further simulations for Comptonizing clouds 
with density profiles $n(r) \propto r^{-p}$ for $p = 1$ and 
$3/2$ and for clouds with temperature profiles of the same
form. Generally, the light curves for a central source can 
be parametrized by a Gamma distribution,
\begin{equation}
f_{int} (t) = A \, t^{\alpha - 1} \, e^{-t / \beta} \, \Theta 
\left(t - {R \over c} \right)
\end{equation}
as suggested by Hua et al. (1997b), where 
\begin{equation}
\Theta (x) = \cases{ 0 & if $x < 0$ \cr
 1 & if $x > 0$ \cr}
\end{equation}
is the Heaviside function and $A$ is a normalization constant.
Inspecting the simulated light curves for the different cases, 
we find that both $\alpha$ and $\beta$ are energy-dependent in both 
the case of a homogeneous cloud as well as a cloud with parameters 
spatially varying. This is in contradiction to the finding of
Hua et al. (1997b) who found a qualitative difference between
the light curves in these two cases. According to our simulations, 
the difference between these two situations is purely quantitative. 
The choice of a much smaller homogeneous Comptonizing region 
compared to the case of an extended cloud with $n(r) \propto r^{-1}$ 
leads to the fact that the differences in the short-term behavior 
($t \ll 1/\beta$) for different energy channels only affects very 
high Fourier frequencies for which no phase and time lags are measured
and which have not been resolved in the simulations of Hua et al. 
(1997b). However, in the situation investigated there, this effect 
only affects very high Fourier frequencies for which no phase and 
time lags are measured anyway. 

Both the early-time slope $\alpha$, describing the light curve 
immediately after the onset of the flare (accounting for the 
light travel time) and the cut-off time $\beta$ (which is
of the order of the photon escape time $t_{esc} \approx 
\tau_T R/c$ for $\tau_T \ugr 1$) tend to increase with 
photon energy. The assumption of a density
gradient (conserving the total Thomson depth $\tau_T$ and 
extent $R$ of the region) leads to a considerable reduction 
of $\alpha$ and a moderate reduction of $\beta$. These changes
are much more pronounced for the high photon energies than
for the lower energy channels, implying that the energy
dependence of $\alpha$ and $\beta$ is reduced due to the
spatial gradient. The assumption of a temperature gradient 
has a similar effect on the light curve parameters than a 
density gradient. In both cases, scattering in the outer 
regions of the cloud becomes less efficient, leading to 
a steeper light curve.

The values of $\alpha$ and $\beta$ yielding a good representation
of the simulated light curves for central soft photon injection 
are listed in Table 1. However, we point out that in the case 
of radial parameter gradients the light curves of the higher 
two energy channels show significant deviations from a 
power-law at the onset of the flare of Comptonized radiation. 
This issue will be discussed further in the next section.

For the case of radiation impinging from outside onto the 
Comptonizing region, we assume that the source of soft
photons is located close to the outer boundary of the
hot plasma region, implying that the radiation enters
the Comptonizing region at angle cosines with respect
to the radial direction which are randomly distributed
between -1 and 0. In this case, the light curve can generally
be parametrized by the sum of a power-law resulting from
single scattering over the light crossing time 
through the source, $2 \, R / c$ plus a multiple
scattering component, described by a light curve of the
form of Eq. (1), i. e.
\begin{equation}
f_{ext} (t) = \Theta (t) \, \left\lbrace A \, \, t^{\alpha - 1} 
\, e^{-t / \beta} + B \, t^{\kappa - 1} \, \Theta \left( 
{2 \, R \over c} - t\right) \right\rbrace,
\end{equation}
where $A$ and $B$ are normalization constants. Figs. 2
and 3 illustrate the energy-dependent light curves for 
scattering of an external soft radiation flare in a 
homogeneous and an inhomogeneous cloud, respectively, 
of the same total radial extent and Thomson depth. The 
resulting Comptonization spectra of these two situations 
are similar to the ones obtained from the situations assumed 
in the case of central photon injection.

For a homogeneous cloud, clearly the first term in Eq. (3)
is dominant, i. e. the parametrization is basically the same
as in the case of a central soft photon source, except in
the two highest photon energy channels, where the 
light curves at $t \ll R/c$ are given by a broken 
power-law with a flat early-time slope. With a more 
pronounced radial density gradient, the second term
in Eq. (3) becomes more important and dominates the shape
of all light curves until $t \approx R/c$ for density
profiles with $p \ugr 1$, where $n(r) \propto r^{-p}$. 
However, although the parametrization
of the light curves for a uniform cloud scattering radiation
impinging from the outside is formally the same as for
scattering of radiation coming from a central soft
photon source, the energy-dependence of the spectral index
$\alpha$ of the light curves for $t \ll R/c$ is significantly
weaker in the case of the external source. This becomes
obvious when comparing Tables 1 and 2, where the parameters 
characterizing the light curves for the two geometrically
different situations are listed, and from inspection of
Fig. 10, where the energy-dependence of the early-time
slope of the light curves is plotted.

It should be noted that the light curves resulting from
our numerical simulations differ significantly from the
analytical results of Lightman and Rybicki (1979b) for
the problem of inverse-Compton reflection off a hot plasma.
This descrepancy is due to the fact that Lightman and Rybicki
used the photon escape probability corresponding to the case
of a semi-infinite, plane-parallel medium with infinite
Thomson depth. Lightman and Rybicki (1997a) have shown 
that the inverse-Compton spectrum in that case is basically 
given by a power-law of photon index 1, which is much harder
than the observed hard X-ray spectrum of Galactic black-hole
candidates (typically $\Gamma \sim 1.5$ -- $2$). Therefore
we have to consider a medium of finite Thomson depth,
$\tau_T \ll 10$.

\section{Fourier transforms, power spectra and time lags}

We performed a numerical Fourier Transform for each energy
band in our simulated light curves and calculated the resulting 
power spectrum densities. Figs. 4 and 5 show the PSD
corresponding to the simulations for external soft photon
injection for the cases $p = 0$ and $p = 1$, respectively,
illustrated in Figs. 2 and 3. Knowing the Fourier transform 
$F_k (\omega) = r_k \, e^{i \phi_k}$ of the light curve in 
energy channel $k$ at Fourier frequency $\omega = 2\pi\nu$, 
the phase difference $\Delta \phi_{kl}$ between two energy 
channels $k$ and $l$ can be computed. The resulting time lag 
between the signals in different energy channels is then given 
by $\Delta t_{kl} = \Delta \phi_{kl} / \omega$. 

The time lags as a function of Fourier period $P = 1 / \nu$ 
resulting from our simulations for central soft photon
injection into a homogeneous and a $p = 1$ cloud, respectively,
are displayed in Figs. 6 and 7. Figs. 8 and 9 show the
respective time lag curves for the case of external soft 
photon injection. In order to suppress numerical
noise contained in the simulated light curves, we
use smooth analytical curves following the light curves
closely for the computation of the hard time lags.
In Figs. 4 and 5, the PSDs from our analytical representations 
(thin curves) are shown in comparison to the PSDs of the 
simulated light curves (thick curves). The figures demonstrate
that the simulated signal of the highest energy channel 
(with the worst photon statistics) is strongly dominated 
by numerical noise at high frequencies. We therefore do
not consider our results concerning the highest energy
channel (31 - 70~keV) in the case of external photon 
injection as fully reliable.

Generally, for both the central and the external soft photon
source, the time lags are increasing with Fourier period 
for frequencies $\omega \ugr \beta_k^{-1} \sim c/(\tau_T R)$ 
in the case of a central soft photon source and for frequencies 
$\omega \ugr \omega_{cr} \equiv c / (2 R)$ for an external 
soft photon source. For lower frequencies (longer periods) 
the respective time lags turn into a flat curve where 
$\Delta\phi_{kl} \propto \omega$. 

For an external soft photon source, no significant deviation
from linearity is found, independent of parameter gradients
or photon energy, except at the lowest photon energies
which are close to the energy of the injected soft 
photons. However, in the case of a central source 
the gradual rise of the light curves at the onset of the 
flare, systematically deviating from a power law, causes 
a flattening of the time lag dependence on the Fourier 
period with increasing photon energies of the respective
channels. In Fig. 6 (central source, $p = 0$), the time 
lag between the lowest two energy channels depends on the 
Fourier period to a power greater than unity, which we 
attribute to the fact that the early-time slopes of the 
two light curves in these energy channels are nearly equal
(see below). 

The effect of the flattening of the hard-time-lag curve with 
increasing photon energy becomes more pronounced with a steeper 
density gradient. For a density $n(r) \propto r^{-3/2}$,
especially the hard lag between the lower energy channels deviates 
from a simple power-law; for the higher channels the curves show 
power-law behaviour with a slope smaller than unity. In this
case, the long-period behavior is no longer flat, but the hard
time lag begins to decline with Fourier period for frequencies
$\omega \ukl \beta^{-1}$.

The power spectra and hard time lags resulting from our simulations 
show a significantly different energy dependence for the two 
different geometries. In the case of the central soft photon 
source and a density or temperature gradient, the PSD is flat 
for low photon energies, while for higher photon energies it 
becomes gradually steeper to higher Fourier frequencies. For the 
$p = 1$ case, this can be approximated by a broken power-law with
break frequency $\omega_{br} \sim \beta^{-1}$. For the solid
sphere, the PSD can generally well be described by a broken
power-law with flat low-frequency slope and the high-frequency
slope increasing with photon energy. 

For an external soft photon source the PSD can roughly be 
represented by a broken power-law for all energy channels 
with break frequency $\omega_{br} \sim c / (2 R)$ and a flat 
low-frequency branch. The power-law slope depends only
very weakly on photon energy and is always close to 2.

Both the analytical representations (1) and (3) of the light curves
in the two geometrical cases considered here can be Fourier-transformed
analytically. The Fourier transforms are
\begin{equation}
F_{int} (\omega) = A { \Gamma (\alpha) \, \beta^{\alpha} \over
\left( 1 + [\beta\omega]^2 \right)^{\alpha/2} } \, e^{i \, \alpha
\, {\rm arctan}(\beta\omega)},
\end{equation}
for the central soft photon source and
\begin{equation}
F_{ext} (\omega) = A {\Gamma (\alpha) \, \beta^{\alpha} \over
\left( 1 + [\beta\omega]^2 \right)^{\alpha/2} } \, e^{i \, \alpha
\, {\rm arctan}(\beta\omega)} + B {\omega_{cr}^{-\kappa}
\over \kappa} \, e^{- i \omega / \omega_{cr}} \, M \left(
1; \, \kappa + 1; \, i {\omega \over \omega_{cr}} \right),
\end{equation}
where $M(a; b; z)$ is Kummer's function, for the external source. 
In the limiting cases $\omega \ll \beta^{-1}, \omega_{cr}$ and
$\omega \gg \beta^{-1}, \omega_{cr}$, respectively, these Fourier 
transforms are
\begin{equation}
F_{int} (\omega) \approx \cases{ A \, \Gamma (\alpha) \, \beta^{\alpha}
\, e^{i \, \alpha \, \beta \omega} & for $\omega \ll \beta^{-1}$ \cr
A \, \Gamma(\alpha) \, \omega^{-\alpha} \, e^{i \, \alpha \, \pi / 2}
& for $\omega \gg \beta^{-1}$ \cr }
\end{equation}
and
\begin{equation}
F_{ext} (\omega) \approx \cases{ A \, \Gamma (\alpha) \, \beta^{\alpha}
\, e^{i \, \alpha \, \beta \omega} + B \, {\omega_{cr}^{-\kappa}
\over \kappa} \, e^{i \, \omega / \omega_{cr}} & for $\omega \ll
\omega_{cr}$ \cr
A \, \Gamma(\alpha) \, \omega^{-\alpha} \, e^{i \, \alpha \, \pi / 2}
+ B \, \Gamma(\kappa) \, \omega^{-\kappa} \, e^{i \, \kappa \pi / 2}
& for $\omega \gg \omega_{cr}$ \cr }
\end{equation}
For the central soft photon source, this yields a Fourier-frequency
independent power spectrum in the low-frequency limit, while for high
frequencies $PSD_{int} \propto \omega^{-2\alpha}$. In general, the hard 
time lags between two energy bands in the high- and low-frequency
limits are
\begin{equation}
\Delta t_{int} = \cases{ 
\Delta(\alpha\beta) & for $\omega \ll \beta^{-1}$  \cr
\Delta\alpha P / 4  & for $\omega \gg \beta^{-1}$, \cr }
\end{equation}
as found by Hua et al. (1997b). If the values of $\alpha$ for two 
energy bands are exactly equal, then it is easy to show that the 
phase difference in the high-frequency limit is $\Delta\phi_{kl} 
\propto \omega^{-1}$ and $\Delta t_{kl} = \alpha \, \Delta\beta \, P^2 / 
(4 \pi^2 \beta_k \beta_l)$. 

From our simulations, we find that the Fourier transforms of the
light curves in the case of an external photon source are dominated
by the second term in Eqs. 5 and 7, respectively, in both limits
$\omega \ll \omega_{cr}$ and $\omega \gg \omega_{cr}$, if the
scattering medium has a strong radial density gradient 
($p \ugr 0.5$). The high-frequency limit of the 
Fourier transform (5) is completely determined by the shape of the 
light curves at $t \ll R/c$ and can always be obtained by simply Fourier 
transforming a function for the form $\tilde f (t) = t^{\mu - 1} 
\, \Theta(t)$, describing the light curve for $t \ll R/c$, without 
accounting for the cutoff at $t \approx 2R/c$ or $t = \beta$, 
respectively. Therefore, for $\omega \gg \omega_{cr}$ we have 
$PSD \propto \omega^{-2 \mu}$ where $\mu = \alpha$ for a weak
density gradient ($p \ukl 0.5$) and low photon energies and 
$\mu = \kappa$ else. 

However, as mentioned in the previous section, this relation
between the early-time slope of the light curve and the high-frequency
branch of the PSD does not apply to the case of the free-fall density 
gradient $p = 3/2$ and internal soft photon injection, where the 
light curves at the onset of the flare significantly deviate 
from a power-law behavior, which leads to a deviation of the PSD 
from a simple power-law at high Fourier frequencies. 

For the hard time lags we find in the case of external soft
photon injection
\begin{equation}
\Delta t_{ext} = \cases{
{1 \over \omega_{cr}}   & for $\omega \ll \omega_{cr}$  \cr
\pi^2 \, P \, \Delta\mu & for $\omega \gg \omega_{cr}$. \cr }
\end{equation}

The qualitative similarity of the asymptotic expressions for the
Fourier transforms for both injection geometries is a consequence 
of the fact that in both cases the rise of the Compton-scattered 
flare is described by a power-law, which results in a power-law
PSD at high frequencies, while for both geometries there is
a fast decay after one light-crossing time $t_c$, which always
results in a constant PSD at frequencies $\omega \ll t_c^{-1}$.

Assuming that the real signal we measure from an X-ray binary 
is composed of multiple shots randomly distributed in time and 
with a certain distribution of shot amplitudes $A_n$, i. e.
\begin{equation}
f(t) = \sum\limits_{n=-\infty}^{\infty} A_n \, f_0(t - t_n)
\end{equation}
where $f_0$ is one of the single-shot light curves (1) or (3),
yields a Fourier transform
\begin{equation}
F(\omega) = F_0 (\omega) \sum\limits_{n=-\infty}^{\infty}
A_n e^{- i \, \omega t_n}
\end{equation}
where $F_0$ is the respective single-shot Fourier transform
(4) or (5). From Eq. (11) it is obvious that the time lags
between different energy bands remain unaffected by the
summation of multiple shots, and in the PSD only near the
average shot repetition frequency some additional structure
is introduced. Therefore the asymptotic expressions (6)
and (7) yield a very good description of the expected
high-frequency branch of the PSD from a model of Comptonized 
random shots of soft photons in the two geometries discussed 
here, and Eqs. (8) and (9) remain valid without modification.

\section{Comparison to observations}

The analysis of the frequency-dependent light curves and
their Fourier transforms enables us to extract important
information from observed or observable properties of the
rapid aperiodic variability of Galactic black-hole candidates.
Observable quantities on which we will focus now are: 
a) the high-frequency power-law index of the PSD and its 
photon-energy dependence, b) the turnover frequency of
the PSD, c) the photon-energy dependence of hard time 
lags with respect to the soft X-rays, d) the turnover
frequency in the hard-time-lag versus Fourier-Period 
relation.

As we have seen in the previous section, the power-law
index of the PSD at high Fourier frequencies is dominated
by the short-term behaviour of individual flares and is
described by the parameter $\alpha$ or $\kappa$ of Eqs.
(1) or (3), respectively. This dependence can be written
as $PSD \propto \omega^{-2\mu}$ where $\mu$ is the
dominant short-time light curve slope ($\alpha$ or 
$\kappa$). In Fig. 10, we have plotted this slope as 
a function of the mean photon energy in the energy bins 
in which the light curves were sampled, for density gradients
of $p = 0$, $1$, and $3/2$, respectively, for both central
and external soft photon injection. 
As we mentioned earlier, in the case of central soft photon 
injection and a density gradient of $p \ukl 1$, a 
significant photon energy dependence results, while for
an external soft photon source the curves are
consistent with a constant slope (recalling that we do
not attribute high significance to our results concerning 
the highest energy channel for external soft photon injection). 

For the external soft photon source, we find on the basis 
of our simulations $\mu_{ext} \approx 1$, irrespective of 
radial parameter gradients. This yields a PSD declining
as $PSD \propto \omega^{-2}$, which is much steeper than
the observed power spectra which are generally flat below
a frequency of $\nu \sim 0.1$~Hz and turn into a power-law
$PSD \propto \nu^{-\Gamma}$. The slope $\Gamma$ has been
measured for Cyg~X-1 to be $\Gamma \approx 1$ 
(\cite{bps74}, \cite{ngm81}) , while for GX~339-4
Motch et al. (1983) found $\Gamma = 1.6 \pm 0.2$, and
Maejimo et al. (1984) evaluated $\Gamma = 1.7$. The
universality of the external-source PSD slope $\alpha
\approx 2$ therefore are marginally consistent with the
value found for GX~339-4, but seems to argue strongly 
against this geometry in the case of Cyg~X-1.

For the case of the central source and a density gradient
$p = 1$, our values of $\alpha$, which are in agreement with 
the ones found by Hua et al. (1997 b), are consistent
with a $PSD \propto \omega^{-1}$ for the energy channel
2.2 --- 4.6~keV. For lower energies, our simulations predict
a slightly flatter PSD, while for higher energies a steeper
PSD is found.

An important observational test of the model of 
Comptonization of centrally injected soft photons
into an inhomogeneous hot medium would be a reliable 
determination of the photon-energy dependence of the 
PSD power-law slope at high Fourier frequencies.

In the case of Cyg~X-1, Nolan et al. (1981) did not 
find any significant difference in the PSD between 
the energy channels 12 --- 30~keV and 30 --- 140~keV, 
while Brinkmann et al. (1974) and Miyamoto \& Kitamoto 
(1989) even found evidence for a flattening of the PSD with 
increasing photon energy, in striking contrast to the 
prediction of the Comptonization model in both geometries 
investigated here.

For GX~339-4, no photon-energy dependence of the PSD
power-law slope was detected (\cite{mrp83}, \cite{mmm84}).

In the case of an external soft photon source, the 
early-time slope $\mu$ of the single-flare light 
curves also determines the hard time lag (cf. Eqs. 
[8] and [9]) at high Fourier frequencies. The
predicted amount of the resulting time lags are
comparable for both scenarios. The deviation from a
power-law behavior of the light curves in the case
of an central soft photon source, however, leads
to a systematic deviation from linearity in the 
dependence of the hard time lag on Fourier period. 
With the parameters used in Hua et al. (1997 b) and
in the energy range considered in Miyamoto et al.
(1988), this curve is well described by a linear
dependence, but for higher photon energy channels,
a significant flattening of this dependence should
be observable if Comptonization of centrally
injected soft photons is responsible for the
rapidly varying hard X-rays. In contrast, such a
deviation from strict linearity should not be
observed in the case of an external soft photon
source.

Very recently, Crary et al. (1997) indeed found a
dependence $\Delta t \propto P^{0.8}$ for the time 
lag of the 50 --- 100~keV photons with respect to
the 20 --- 50~keV signal from Cyg~X-1. This seems
to be another hint toward the central soft photon 
injection scenario. Also the observed approximate
dependence of the $\Delta t_{kl} \propto \ln (E_2 / E_1)$
(\cite{mkk88}) is more consistent with this 
geometry (see Fig. 10 and recall that approximately 
$\Delta t \propto \Delta \alpha$ for central injection).

The available data on the phase lags in GX~339-4 and 
GS~2023+338 (Miyamoto et al. 1992) do not show an 
obvious trend and definitely need to be supplemented 
by more sensitive measurements comparing a larger 
number of energy channels in order to test the 
Comptonization models investigated in this paper.

In the discussion of Section 3, we have seen that the 
break frequencies defining the transition between
a flat PSD and a steep power-law as well as the transition
between the linear dependence of the hard time lag on 
Fourier period to a constant time lag are both expected
around
\begin{equation}
\nu_b = \cases{ 
c / (2 \, \pi \, \tau_T \, R) & for central injection \cr
c / (4 \, \pi \, R)           & for external injection \cr }
\end{equation}
if $\tau_T \ugr 1$. Since for most Galactic black-hole 
candidates a Thomson depth of 1 -- a few is appropriate 
to model the hard X-ray spectrum via Comptonization of soft
photons, Eq. (12) yields roughly the same size estimate 
for both injection geometries,
\begin{equation}
R \sim \cases{ 5 \cdot 10^{10} \, \left( \nu_{b, -1} \, \tau_T 
\right)^{-1} \> {\rm cm} & for central injection \cr
2.5 \cdot 10^{10} \, \nu_{b, -1}^{-1} \> {\rm cm} & for external
injection \cr}
\end{equation}
where $\nu_{b, -1}$ is the break frequency in units of 0.1~Hz.
Since the observed flattening frequencies of the PSDs
of GBHCs are of the order $\nu_{flat} \sim 0.1$~Hz
and the measurements of Miyamoto et al. (1988, 1993) yield a
lower limit on the turnover Fourier period of the hard time 
lag in the same frequency range, Eq. (16) reveals a major 
problem of the Comptonization scenario as an explanation 
for the rapid aperiodic variability of Galactic black-hole 
candidates in general. If indeed the spectral breaks in 
the PSD and the hard time lags are the result of inverse 
Compton scattering as analyzed in the previous sections, 
then the estimated size of the Comptonizing region is much larger 
than the expected extent of an accretion-disk corona. This 
problem is obviously even more severe for the case of external
photons impinging on a hot inner-disk torus.

The problem resulting from the break in the power spectrum could
be overcome assuming a soft radiation flare of much longer intrinsic
duration than the light crossing time of the inner disk region.
In this case, the turnover frequency of the PSD would yield an estimate
for the typical intrinsic duration of the flare, $\nu_{flare}^{-1}
\sim$~1---10~s, which, e. g., is of the order of the radial drift 
time scale in the inner accretion disk. This would generally
produce an additional break in the PSD at the turnover frequency
of the single-shot power spectrum. A second break or a smooth
steepening of the PSD around $\sim 10$~Hz has indeed been observed 
in the low states of several GBHCs.

However, we expect that the assumption of a broad flare as opposed
to a $\delta$-shot in time does not shift the turnover of the 
hard-time-lag versus Fourier-period curves if it is indeed due 
to inverse Compton scattering. Therefore, the Fourier-period
dependence of the hard time lag is still a puzzle in the framework
of Comptonization models. As one possible solution to this problem
we suggest an intrinsic hardening of the soft input photon spectrum 
during the evolution of an extended flare, which could be caused
by a hot spot in the accretion disk heating up as it spirals inward.
This idea will be worked out in detail in a forthcoming paper.

\section{Summary and conclusions}

We presented a systematic study on the predicted Fourier 
power spectra and hard time lags resulting from two
fundamentally different geometries in Comptonization models for
the hard X-ray emission of Galactic black-hole candidates: a
flaring soft photon source located in the center of the Comptonizing
region (e. g., a hot corona) and a soft photon source located 
outside a hot inner disk region. The different scenarios
yield different predictions for the Fourier power spectra
and fundamental differences in the photon-energy dependence
of the Fourier PSD slope and of the slope describing the 
hard-time-lag versus Fourier-period relation. 

We found that Comptonization of flares of external soft 
radiation generally leads to a weak photon-energy dependence
of the power spectrum and the hard time lag, with the power
spectrum being marginally consistent with the PSD measured
for GX~339-1, but significantly steeper than for Cyg~X-1.
Comptonization of centrally injected soft photon flares,
in turn, leads to a significant photon-energy dependence
in the PSD power-law slope at high Fourier frequencies and
photon-energy dependent deviations from linearity of the
hard time lags as a function of Fourier period. With a
density gradient $n(r) \propto r^{-1}$, the simulated PSD
for medium-energy X-rays is consistent with the slope 
observed in Cyg~X-1.

The turnover in the Fourier-frequency dependent hard 
time lag curve poses a severe challenge to Comptonization 
models for the rapid aperiodic variability in GBHCs in general 
due to the resulting large size estimates.

Detailed measurements of the energy-dependent properties 
of the rapid aperiodic GBHC X-ray variability are strongly
encouraged. Using the diagnostic tools developed in this
paper, they can shed light on the question of the geometry
of black-hole accretion disks and the location of the
Comptonizing region and may serve to rule out a specific 
geometry or even the Comptonization models as a mechanism 
involved in the rapid aperiodic X-ray variability of 
Galactic black-hole candidates in general.

\acknowledgements{This work is partially supported by NASA
grant NAG~5-4055.}

\begin{figure}
\rotate[r]{
\epsfysize=12cm
\epsffile[150 0 550 500]{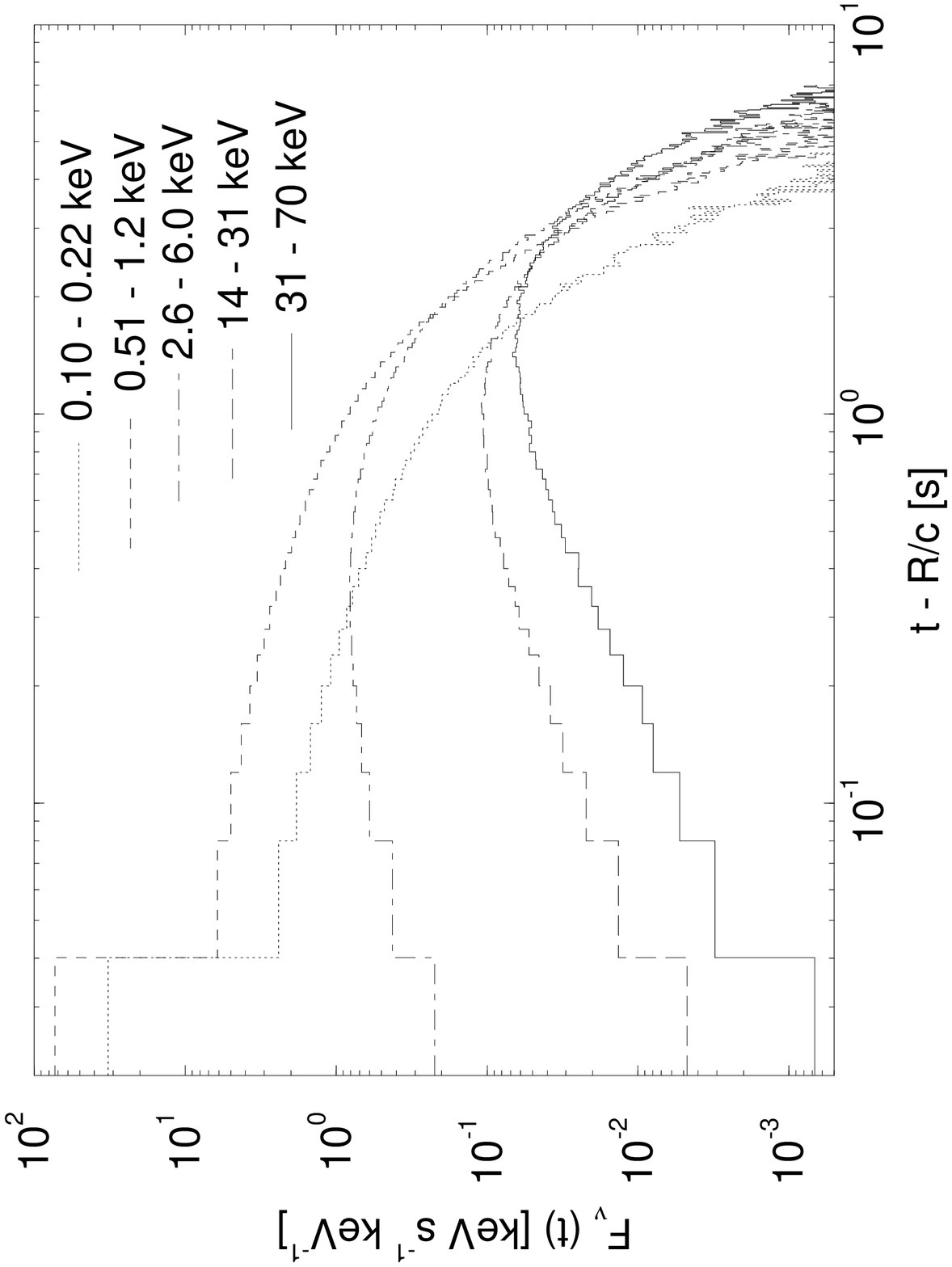}
}
\caption[]{Energy-dependent light curves for Comptonization of
an instantaneous soft radiation flare at $t = 0$ in the center
of the Comptonizing region with uniform density ($p = 0$) and 
Thomson depth $\tau_T = 1$, $R = 3 \cdot 10^{10}$~cm, $kT_e =
100$~keV}
\end{figure}

\begin{figure}
\rotate[r]{
\epsfysize=12cm
\epsffile[150 0 550 500]{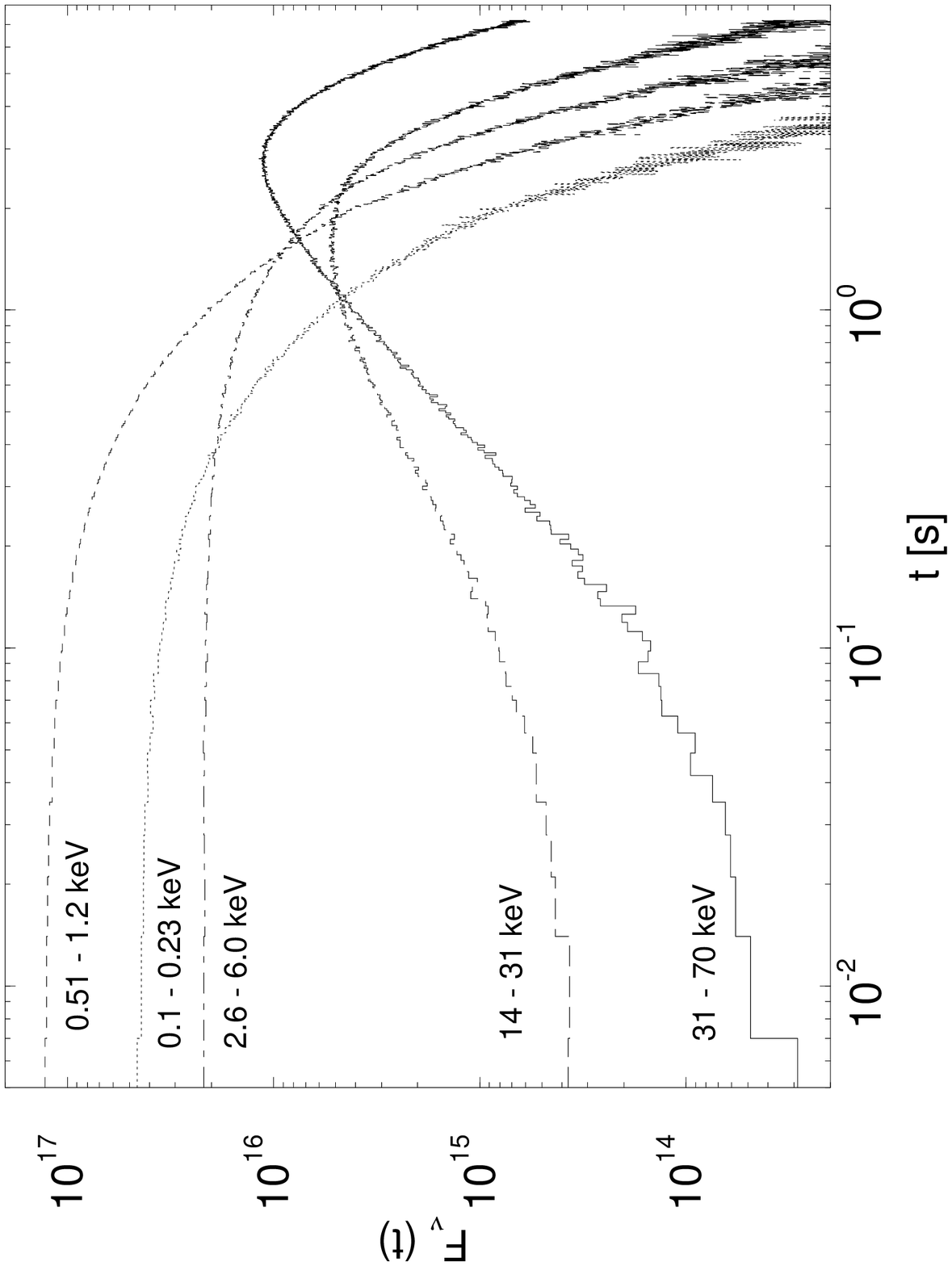}
}
\caption[]{Energy-dependent light curves for Comptonization of
an instantaneous soft radiation flare at $t = 0$ outside a
homogeneous Comptonizing region ($p = 0$) with $\tau_T = 3$, 
$R = 3 \cdot 10^{10}$~cm, $kT_e = 100$~keV}
\end{figure}

\begin{figure}
\rotate[r]{
\epsfysize=12cm
\epsffile[150 0 550 500]{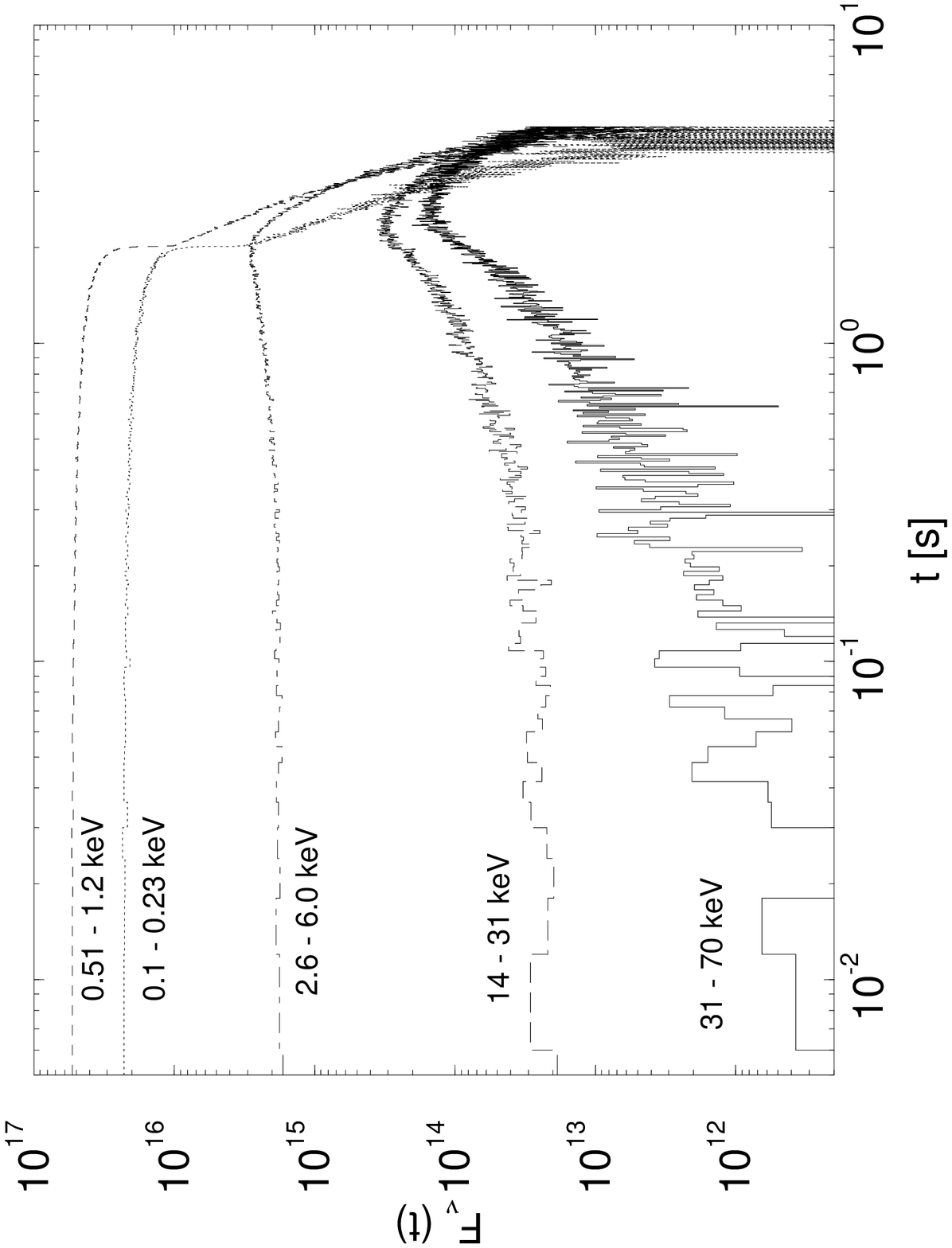}
}
\caption[]{Energy-dependent light curves for Comptonization of
an instantaneous soft radiation flare at $t = 0$ outside a
homogeneous Comptonizing region with density gradient $p = 1$ 
with $\tau_T = 2.1$, $R = 3 \cdot 10^{10}$~cm, $kT_e = 100$~keV}
\end{figure}

\begin{figure}
\rotate[r]{
\epsfysize=12cm
\epsffile[150 0 550 500]{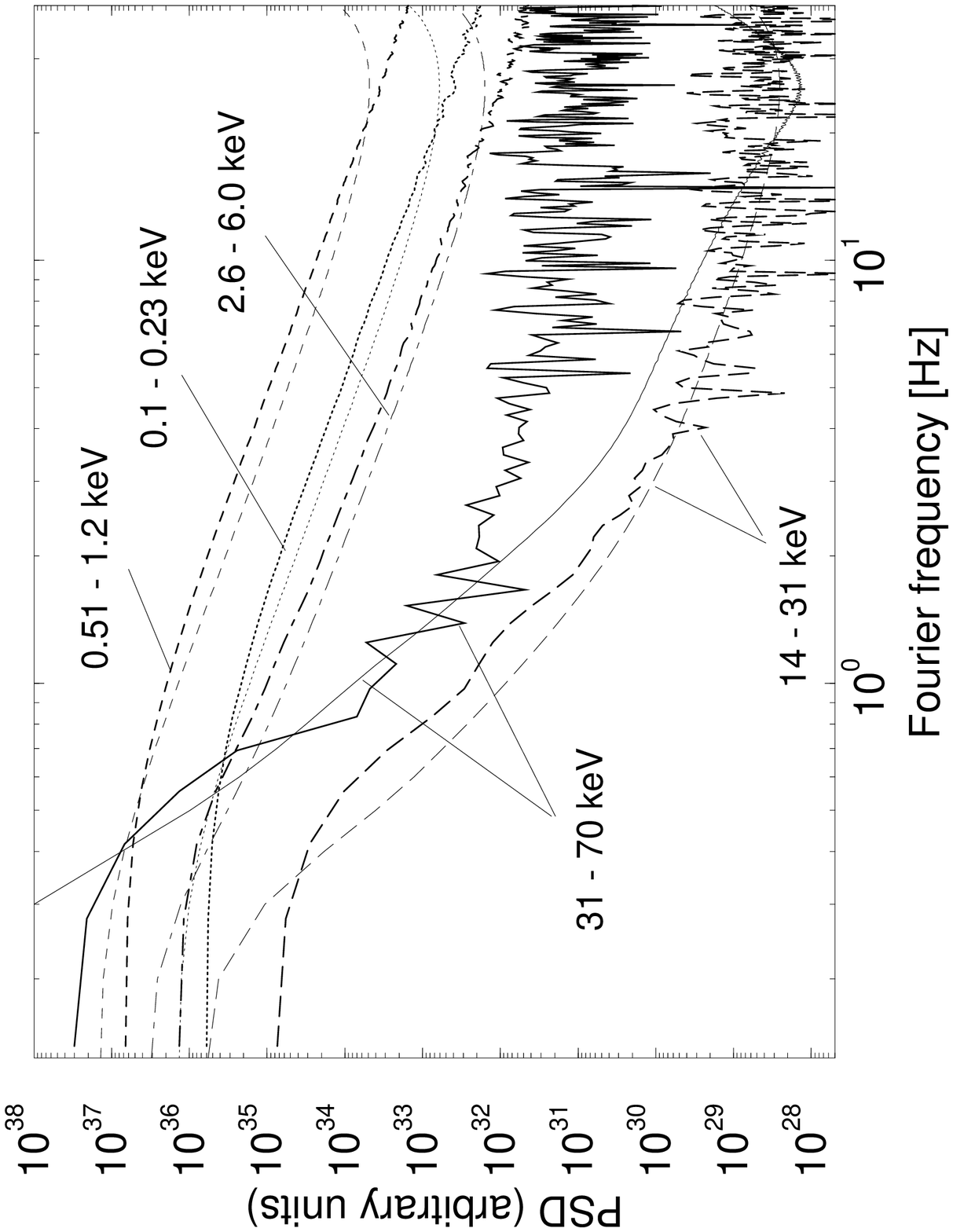}
}
\caption[]{Energy-dependent power spectrum densities for external
soft photon injection, $p = 0$, parameters see Fig. 2 (thick curves:
Fourier transform of simulated light curves; thin curves: Fourier
transform of analytical representation)}
\end{figure}

\begin{figure}
\rotate[r]{
\epsfysize=12cm
\epsffile[150 0 550 500]{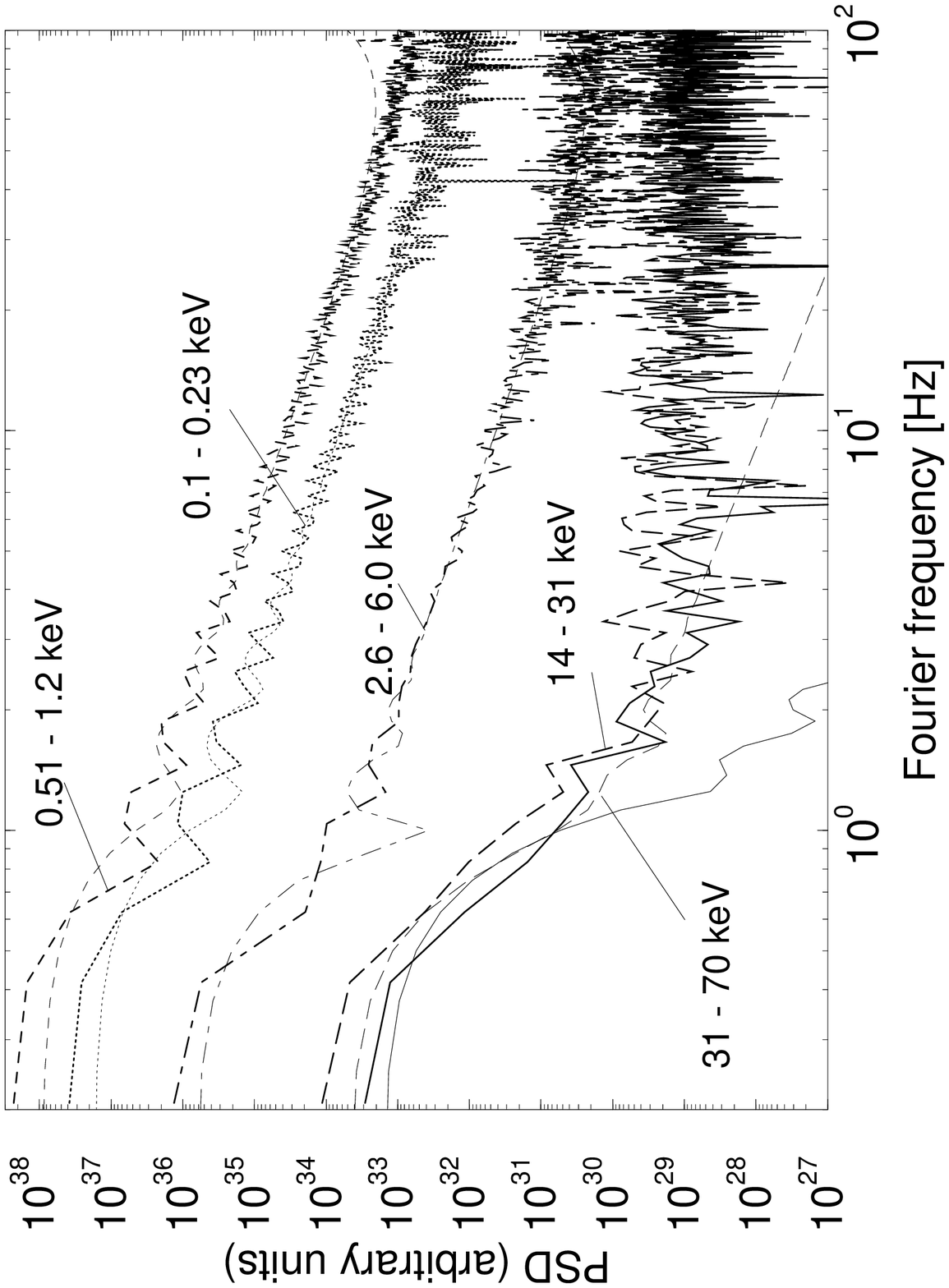}
}
\caption[]{Energy-dependent power spectrum densities for external
soft photon injection, $p = 0$, parameters see Fig. 3 (thick curves:
Fourier transform of simulated light curves; thin curves: Fourier
transform of analytical representation)}
\end{figure}

\begin{figure}
\rotate[r]{
\epsfysize=12cm
\epsffile[150 0 550 500]{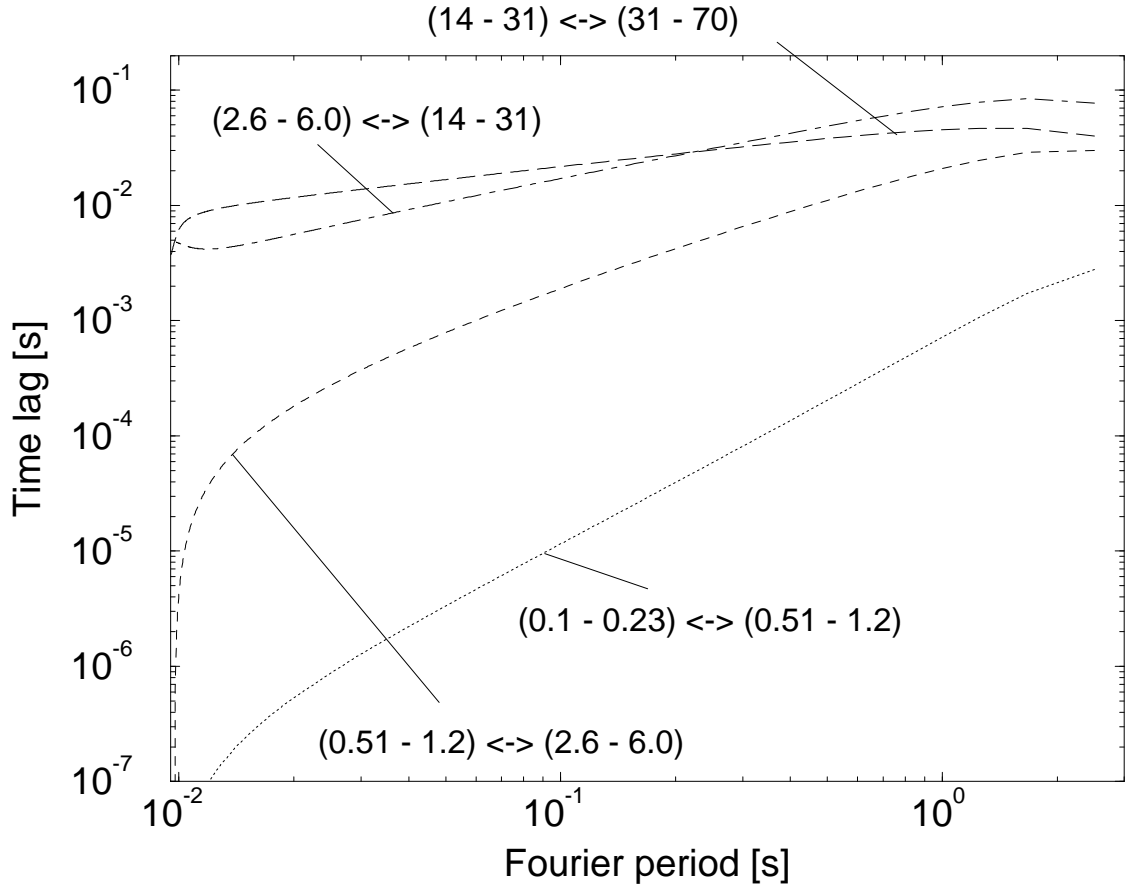}
}
\caption[]{Fourier-frequency-dependent hard time lags for 
internal soft photon injection, $p = 0$; parameters see Fig. 1}
\end{figure}

\begin{figure}
\rotate[r]{
\epsfysize=12cm
\epsffile[150 0 550 500]{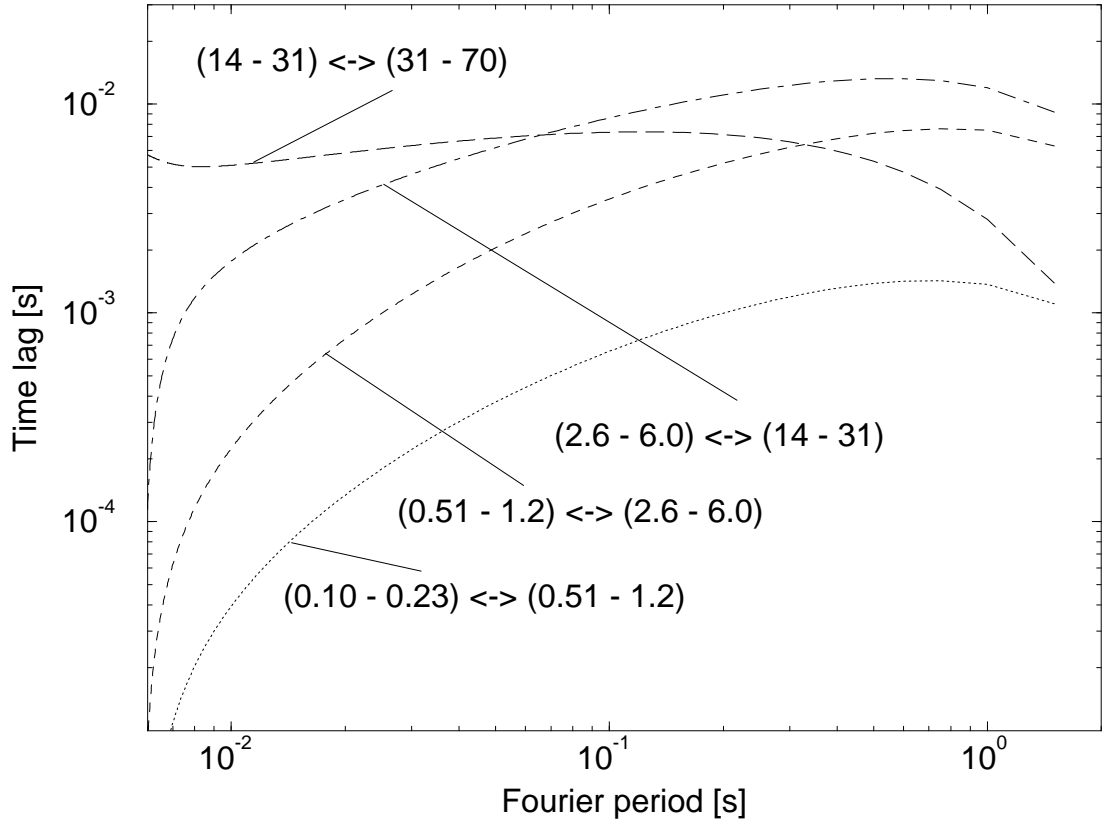}
}
\caption[]{Fourier-frequency-dependent hard time lags for 
internal soft photon injection, $p = 1$, $\tau_T = 0.7$,
$R = 3 \cdot 10^{10}$~cm, $kT_e = 100$~keV}
\end{figure}

\begin{figure}
\rotate[r]{
\epsfysize=12cm
\epsffile[150 0 550 500]{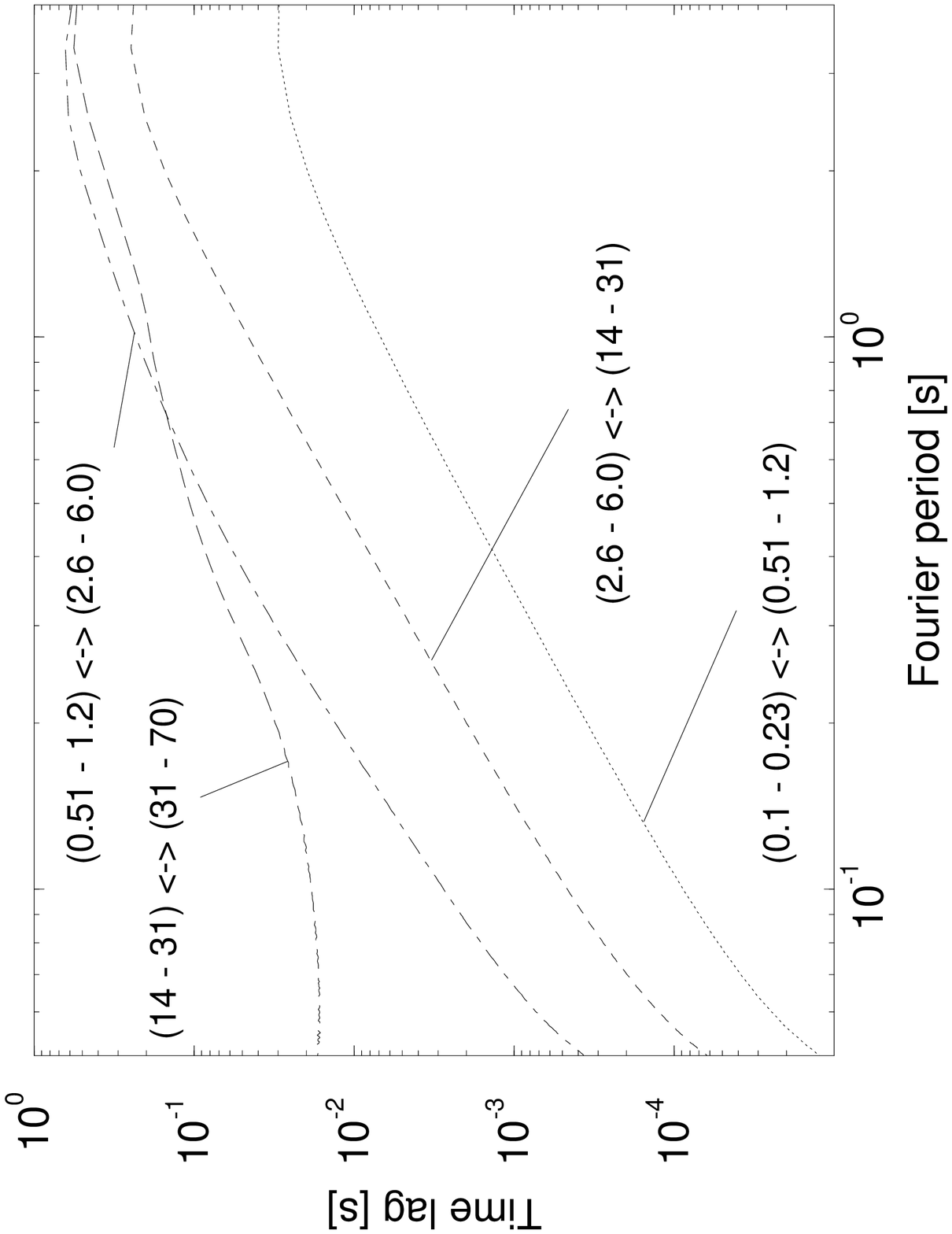}
}
\caption[]{Fourier-frequency-dependent hard time lags for 
external soft photon injection, $p = 0$; parameters see
Fig. 2. The irregular structures around 0.5 -- 1 s are due 
to numerical noise}
\end{figure}

\begin{figure}
\rotate[r]{
\epsfysize=12cm
\epsffile[150 0 550 500]{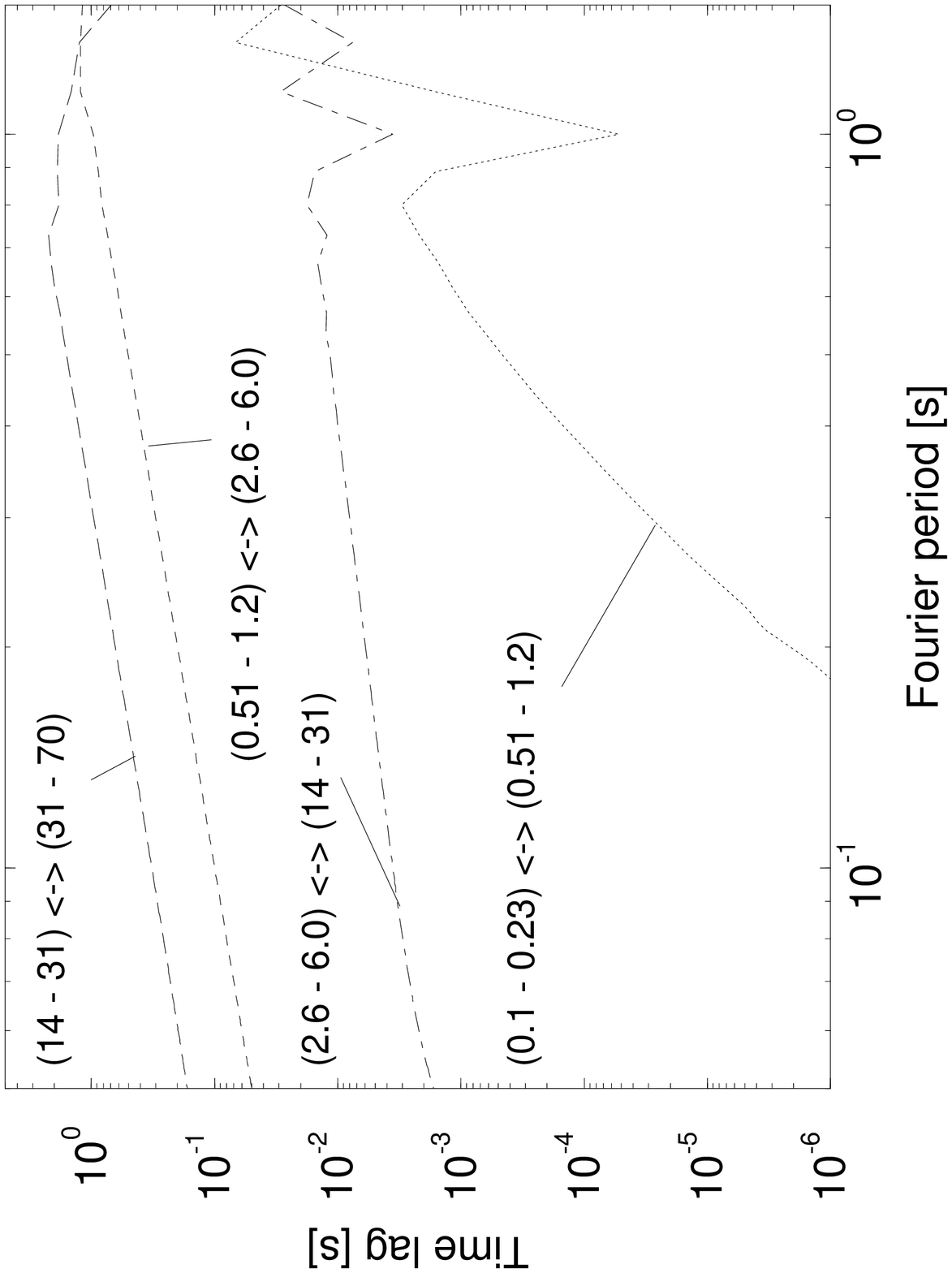}
}
\caption[]{Fourier-frequency-dependent hard time lags for 
external soft photon injection, $p = 1$; parameters see
Fig. 3. The irregular structures around 0.5 -- 1 s are due 
to numerical noise}
\end{figure}

\begin{figure}
\rotate[r]{
\epsfysize=12cm
\epsffile[150 0 550 500]{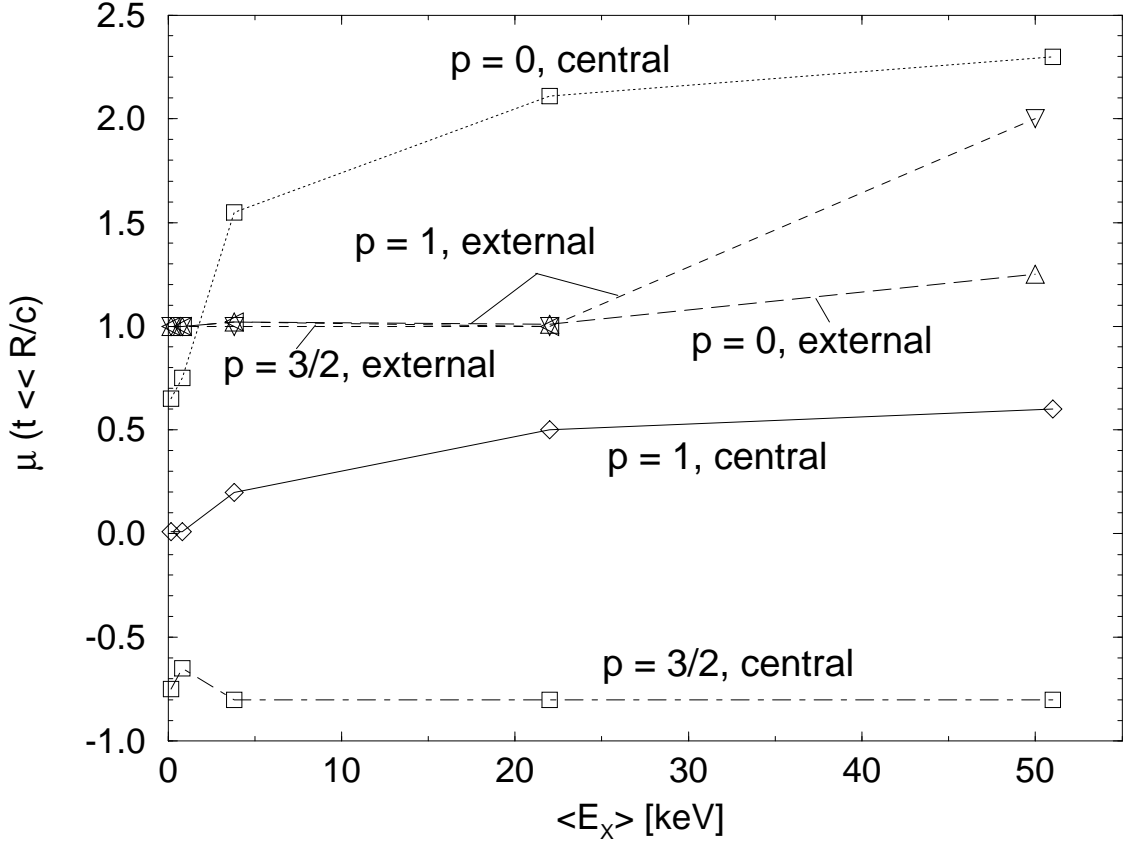}
}
\caption[]{Early-time light curve slopes $\mu$ ($= \alpha$ or 
$\kappa$, respectively; cf. Sect. 2) as a function of the mean 
photon energy in the different energy bins for the cases $p = 0$,
$1$, and $3/2$, for internal and external soft photon injection. 
The curves are labeled by the slope density gradient slope $p$ 
and the location of the soft photon source.
}
\end{figure}

\newpage

\begin{deluxetable}{cccc}
\tablewidth{9cm}
\tablecaption{Lightcurve fit parameters for Comptonization of a flare
of soft radiation in the center of the scattering cloud. For definition
of the parameters see Eq. (1).}
\tablehead{
\colhead{p} & \colhead{energy channel [keV]} & 
\colhead{$\alpha$} & \colhead{$\beta$ [s]}}
\startdata
0   & 0.10 -- 0.22       & 0.65     & 0.67        \nl
0   & 0.51 -- 1.2        & 0.75     & 0.76        \nl
0   & 2.6 -- 6.0         & 1.55     & 0.65        \nl
0   & 14 -- 31           & 2.11     & 0.85        \nl
0   & 31 -- 70           & 2.30     & 1.05        \nl
\hline
1   & 0.10 -- 0.22       & 0.0      & 0.77        \nl
1   & 0.46 -- 1.0        & 0.0      & 0.71        \nl
1   & 2.2 -- 4.6         & 0.2      & 0.69        \nl
1   & 10 -- 22           & 0.5      & 0.63        \nl
1   & 46 -- 100          & 0.6      & 0.63        \nl
\hline
1.5 & 0.10 -- 0.22       & - 0.75   & 0.90        \nl
1.5 & 0.51 -- 1.2        & - 0.65   & 0.90        \nl
1.5 & 2.6 -- 6.0         & - 0.80   & 0.90        \nl
1.5 & 14 -- 31           & - 0.80   & 0.90        \nl
1.5 & 31 -- 70           & - 0.80   & 0.80        \nl
\enddata
\end{deluxetable}

\begin{deluxetable}{cccccc}
\tablewidth{12cm}
\tablecaption{Light curve fit parameters for Comptonization of a flare
of soft radiation impinging from outside the scattering cloud. For 
definition of the parameters see Eq. (3).}
\tablehead{
\colhead{p}        & \colhead{energy channel [keV]} & 
\colhead{$\alpha$} & \colhead{$\beta$ [s]}          & 
\colhead{$\kappa$} & \colhead{A/B}}
\startdata
0   & 0.10 -- 0.22         & 1.00     & 0.48        & n. det.  & $\infty$ \nl
0   & 0.51 -- 1.2          & 1.00     & 0.54        & n. det.  & $\infty$ \nl
0   & 2.6 -- 6.0           & 1.02     & 1.55        & n. det.  & $\infty$ \nl
0   & 14 -- 31             & 2.28     & 1.30        & 1.01     & 25.0     \nl
0   & 31 -- 70             & 2.90     & 2.00        & 1.25     & 44.8     \nl
\hline
1   & 0.10 -- 0.22         & 8.00     & 0.21        & 1.00     & 9.09     \nl
1   & 0.51 -- 1.2          & 8.00     & 0.21        & 1.00     & 13.5     \nl
1   & 2.6 -- 6.0           & 8.00     & 0.25        & 1.00     & 33.3     \nl
1   & 14 -- 31             & 8.50     & 0.29        & 1.00     & 60.0     \nl
1   & 31 -- 70             & 9.00     & 0.32        & 2.00     & 20.0     \nl
\hline
1.5 & 0.10 -- 0.22         & 4.00     & 0.25        & 1.00     & 4.29     \nl
1.5 & 0.51 -- 1.2          & 4.00     & 0.28        & 1.00     & 11.3     \nl
1.5 & 2.6 -- 6.0           & 8.00     & 0.27        & 1.02     & 13.0     \nl
1.5 & 14 -- 31             & 8.00     & 0.28        & 1.00     & 14.3     \nl
1.5 & 31 -- 70             & n. det.  & n. det.     & n. det.  & n. det.  \nl
\enddata
\end{deluxetable}

\end{document}